\documentclass[aps,nofootinbib,preprintnumbers,citeautoscript]{revtex4}
\usepackage{amsmath,amssymb}
\usepackage{enumerate}
\usepackage{graphicx}
\usepackage{amsthm}
 \usepackage{mathtools}
 \usepackage{textcomp}
  \usepackage{braket}
    
\begin{document}
\title{Matrix Product State Based Quantum Classifier}

\author{Amandeep Singh Bhatia$^\ast$, Mandeep Kaur Saggi, Ajay Kumar, Sushma Jain  \\
\textit{Department of Computer Science, Thapar Institute of Engineering \& Technology, India} \\
E-mail: $^\ast$amandeepbhatia.singh@gmail.com}

	\begin{abstract}
	In recent years, interest in expressing the success of neural networks to the quantum computing has increased significantly. Tensor network theory has become increasingly popular and widely used to simulate strongly entangled correlated systems. Matrix product state (MPS) is the well-designed class of tensor network states, which plays an important role in processing of quantum information. In this paper, we have shown that matrix product state as one-dimensional array of tensors can be used to classify classical and quantum data. We have performed binary classification of classical machine learning dataset Iris encoded in a quantum state. Further, we have investigated the performance by considering different parameters on the ibmqx4 quantum computer and proved that MPS circuits can be used to attain better accuracy. Further, the learning ability of MPS quantum classifier is tested to classify evapotranspiration (ET$_{o}$) for Patiala meteorological station located in Northern Punjab (India), using three years of historical dataset (Agri). Furthermore, we have used different performance metrics of classification to measure its capability. Finally, the results are plotted and degree of correspondence among values of each sample is shown.
\end{abstract}
\maketitle



\theoremstyle{plain}
\newtheorem{thm}{Theorem}[section]

\theoremstyle{definition}
\newtheorem{defn}{Definition}[section]
\newtheorem{exmp}{Example}[section]

\section{Introduction}
Quantum computing is a winsome field that deals with theoretical computational systems (i.e., quantum computers) combining visionary ideas of Computer Science, Physics, and Mathematics. It concerns with the behaviour and nature of energy at the quantum level to improve the efficiency of computations. Feynman \cite{feynman1982simulating} initially proposed the idea of quantum computing in 1982 after performing a quantum mechanics simulation on a classical computer. Quantum computing relies upon the quantum phenomena of entanglement, superposition and interference to perform operations, which are generally considered as resources for this speed up. Although, the full influence of quantum computing is probably more than a decade away. But, it has the potential to transform the information processing and promises a wide range of applications in the area of quantum chemistry, high energy physics and condensed matter, which are not tractable on classical computers.

In the last decade, the simulation of open and closed quantum systems has got overwhelming response. The study of tensor network theory taking a central role in quantum physics and beyond. It is simply a countable group of tensors associated by contractions. Tensor network states are a new language, based on entanglement, for quantum many-body systems \cite{gao2017efficient}. Tensor network states are classified on the basis of dimensions along which the tensors are traversed. It is widely used to simulate strongly entangled correlated systems and to represent quantum states and circuits \cite{schuld2018circuit}. ‘Tensor network methods’ is the term associated with the tools, which are widely employed in experimental and quantum theoretical applications of machine learning. The matrix product state (MPS) is the most prominent example of tensor network states which is maximally unbalanced. It can be observed by the maximum entanglement entropy without even forfeiting one-dimensional distributions expressiveness. Matrix product tensor networks has the ability to  surround the whole input or output state space. By using classical resources, tensor networks have shown impressive results for supervised and unsupervised learning tasks \cite{pestun2017tensor}. For large dimensions, tree tensor network (TTN) and multi-scale entanglement renormalization ansatz (MERA) have shown to be equivalent with neural networks. These methods have seen many breakthroughs and transformations to different domains of physics, mathematics and computer science.

Matrix product states are compelling for their wide range of practical applications: supervised learning \cite{miles2016supervised}, quantum dynamics \cite{bhatia2018neurocomputing}, simulating quantum computation \cite{gg}, quantum finite state machines \cite{bhatia2018quantifying}, unsupervised learning \cite{han2018unsupervised}, simulating MPS on a quantum computer \cite{bhatia2018simulation}, quantum machine learning MPS \cite{biamonte2018quantum} and many more. Matrix product states are complete, where low entangled states are represented efficiently, which is not possible with large dimensions tensor network states. It can be also employed in various emerging technologies such as quantum cryptography, optical computing, dynamics quantum clustering and image recognition. It obeys one-dimensional area law, dense in nature, finitely correlated tensors, translational invariance and suited for describing the higher-dimensional systems also. Recently, MPS method have been introduced to compress the weights of neural network layers and classify the images.

Through the effective deployment of Information and Communication Technologies (ICTs), India's agriculture sector has been transformed from the traditional to modern practice. The combination of ICTs and analytics can provide novel ways and ideas to do socially accepted and profitable agriculture. It will be also beneficial for the environment (e.g., soil, water, climate). In future, there is a need of threefold innovation: technological, economical and social agriculture. As the technology rapidly spreads in few decades, big data analytics and quantum machine learning are the keys to fostering a new revolution in agriculture. It has evolved technology to solve real-world problems based on historical data, machine-generated data and real-time streaming data.
Quantum computing can also bring revolution through its ability to handle experimental data, which can produce numerous solutions in various areas such as healthcare, smart city, smart agriculture etc \cite{saggi2018survey}. In developing countries, the farmers with skill level, limited knowledge and adoption of new expensive technologies are the significant factors.  There are some tough questions that need to be addressed. Can farmers manage the farm, grow and harvest the crops simultaneously.? Can they build smart network which can be used in field-testing of next generation of quantum Computers?. Quantum computers promise to revolutionize the complicated data processing problems. The future of quantum computers promise us a tool that can handle more data efficiently and can surpass any conventional system. Quantum machine learning techniques are also closely tied with a variety of application areas. In this paper, we have considered the supervised machine learning task of classifying the Iris and climatic dataset on quantum computer using matrix product state quantum classifier. The organization of paper is as follows. Section 2 is devoted to related work. In Section 3, encoding of data and qubit efficient MPS classifier is described. In section 4, we have shown the model development and experimental results. Finally, Section 5 is the conclusion.

\section{Prior work}
The combination of neural networks and classical machine learning models with the efficiency of quantum computing have experienced incredible responses in the last few years. Initially, Harrow et al. \cite{harrow2009quantum} designed a quantum algorithm to approximate the features of solution of a set of \textit{N} linear equations, which runs in polynomial time. On comparing with the best classical algorithm, it is exponentially faster. In 2014, Rebentrost et al. \cite{rebentrost2014quantum} presented a quantum support vector machine that can be implemented on quantum computer with complexity $O(log NM)$ logarithmic in size of training and classification stages. It has been observed that in contrast to classical algorithms, an exponential speedup is gain.

The accurate estimation of evapotranspiration (ET$_{o}$) is a crucial issue for agriculture planning because it plays a significant role in irrigation water scheduling for using water efficiently. Evapotranspiration is a vital component of the hydrological cycle and there are a large number of alternative models for representing ET$_{o}$ processes. It can be measured directly by experimental techniques, e.g. eddy covariance systems, lysimeters and Bowen ratio energy balance \cite{kool2014review, marti2015modeling}, but these methods are complex and not available in many regions \cite{allen1998crop} due to high cost. Therefore, development of mathematical models for ET$_{o}$ estimation is highly needed. To further support a range of ET$_{o}$ modeling studies, there is a need to facilitate the implementation of different ET models in a convenient, consistent and efficient manner. There are some software packages focusing on specific ET$_{o}$ modeling needs and aspects. Recently, Saggi and Jain developed H$_{2}$O model framework based on deep learning for predicting ET$_{o}$ of Patiala and Hoshiarpur stations \cite{saggi2019reference}.

In 2017, Otterbach et al. \cite{otterbach2017unsupervised} investigated a hybrid quantum algorithm for unsupervised learning task known as clustering on 19-qubit quantum computer. It has been shown that noise is enabled by using gradient-free Bayesian optimization and offers an optimal solution with high probability for all random problem instances. Recently, Farhi and Neven \cite{farhi2018classification} introduced the concept of classification with quantum neural network (QNN). It has presented general framework for supervised learning to represent labeled classical or quantum data. It consists sequence of unitary transformations on the input encoded quantum state and Pauli operator is measured on final output qubit. Further, Farhi and Neven \cite{farhi2018classification}  considered a real world data containing sampled images of two distinct sets of handwritten digits. It has shown the learning capability of QNN to exactly determine the two data sets.

Initially, Miles and Schwab \cite{miles2016supervised} proposed a quantum-inspired tensor networks based framework for multi-class supervised learning. The matrix product states based model is used for classifying the images (MNIST) dataset and acquired less than 0.01 testing error. Han et al. \cite{han2018unsupervised} introduced a MPS based generative modeling for unsupervised learning tasks. They have considered Bars and Stripes random
binary patterns and the MNIST handwritten digits to investigate the features, abilities and shortcomings. It has been shown that MPS exhibits much stronger ability of learning as compared to inverse Ising model and Hopfield model. Liu et al. \cite{09111} implemented entanglement-guided architectures to classify images, where quantum states are written in MPS. In 2018, Liu et al. \cite{liu2017machine} proposed two-dimensional training tree-like tensor networks as classifiers for image recognition problems and tested on MNIST and CIFAR datasets. It has been shown that the proposed algorithm encodes classes of images into a tensor network state optimally and can be characterized by quantum entanglement. Glasser et al. \cite{glasser2018neural} introduced generalized tensor networks and discussed the relationship between restricted Boltzmann machines and String-Bond States. It has been shown that  generalized tensor networks can be associated exactly and can classify images accurately with smaller bond dimension.

Gardas et al. \cite{gardas2018quantum} simulated many body quantum systems using a hybrid classical-quantum
algorithm, where wave function of quantum Ising model is represented using Boltzmann machine. Further, the neural network is trained using D-wave quantum annealer and the ground state energy is calculated. 
Huggins et al. \cite{huggins2018towards} proposed tensor network based quantum computing approaches for generative and discriminative tasks. The main purpose is to generate samples from a probability distribution and assign labels to images. The experimentation is executed on quantum hardware using optimization procedure for handwritten classes of images and noise resilience is tested of the training model. Grant et al. \cite{gg} introduced the concept of hierarchical quantum classifiers and executed binary classification for classical and quantum data. Two classical machine learning datasets Iris and MNIST are considered and deployed the classifiers on quantum computer. It has shown impressive results and better accuracy by considering different unitary parameters. In this paper, following contributions are claimed: 
\begin{itemize}
\item  We demonstrate that matrix product state as one-dimensional array of tensors can be used to classify quantum mechanical data in addition to classical dataset.
\item We encode classical dataset (Iris and Agri) into quantum entangled state, which is given as an input to MPS tensor network quantum circuit. 
\item Four and six qubit inputs are taken for Iris and Agri dataset and measurement is performed on quantum circuit.
\item To investigate the performance, MPS classifier on real-time quantum device (ibmqx4) is deployed. 
\end{itemize}
\section{Matrix product state}
Matrix product state concedes the extent of entanglement in bond dimensions. It is a method of tensor network, where the tensors are connected in a one-dimensional geometry. Figure 1 shows the MPS as one-dimensional array of tensors and an instance of finite system of 5 sites. It provides an efficient approximation of realistic local Hamiltonians and can be produced sequentially by tensors. In fact, any pure quantum state can be described by substituting the coefficients e.g. rank-\textit{N} tensor by \textit{N}-rank 3 tensors and 2-rank by 2 tensors. In MPS, a pure quantum state $\ket{\phi}$ is represented as:
\begin{equation}
\ket{\phi}=\sum_{\sigma_1, \sigma_2,...\sigma_L}^{d} Tr[M_1^{\sigma_1} M_2^{\sigma_2}...M_L^{\sigma_L}] \ket{\sigma_1, \sigma_2,...\sigma_L}
\end{equation}
where $M_i^{\sigma_i}$ are complex square matrices, \textit{d} is dimension, $\sigma_i$ represents the indices i.e. \{0, 1\} for qubits and \textit{Tr}() denotes trace of matrices \cite{bhatia2018quantifying}.  

\begin{figure}[h]
\centering
\includegraphics[scale=0.5]{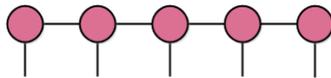}
\caption{Representation of MPS with 5 sites} \label{1}
\end{figure}

\subsection{Encoding of classical data}
In quantum mechanics, the \textit{N} independent systems can be combine by performing tensor product operation on their respective state vectors \cite{miles2016supervised, huggins2018towards}. Consider a feature map
\begin{equation}
\phi^d(x)=\phi^{s_1}(x_1) \otimes \phi^{s_2}(x_2) \otimes...\otimes \phi^{s_N}(x_N)
\end{equation}
where $s_j$ are indices run over the local dimension \textit{d} such that \textit{d}=$\{s_1,s_2,...,s_N\}$. Therefore, each state vector $x_j$ is mapped to full feature map $\phi(x)$ in a \textit{d}-dimensional space. Fig 2 shows the tensor diagram of full feature map $\phi(x)$. 

\begin{figure*}[!h]
\centering
\includegraphics[scale=0.5]{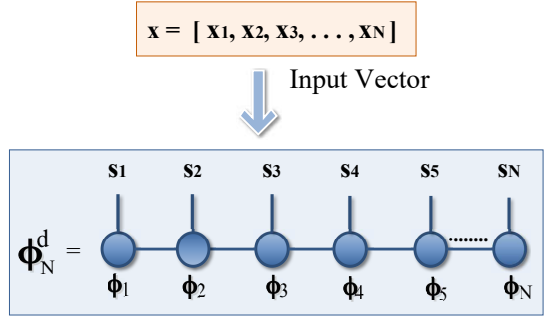}\hfil
\caption{Mapping of input vector to order \textit{N} tensor}
\end{figure*}
Before illustrating the MPS tensor network, it is very crucial to encode classical machine learning dataset into quantum state. Consider a classical dataset $S=\{(x^d, y^d)\}^{D}_{d=1}$ for binary classification, where $y^d \in \{0,1\}$ are class labels for \textit{N}-dimensional input vectors such that $x^d \in$ ${\rm I\!R}^N$. We have normalized the input vectors to lie in $[-\pi, \pi]$.  Thus, the qubit $\phi$ is represented as
\begin{equation}
\phi^d_n=cos(x^d_n)\ket{0}+sin(x^d_n)\ket{1}
\end{equation}
\begin{equation}
\phi^d_n=\begin{bmatrix}
cos(x^d_1) \\
sin(x^d_1)
\end{bmatrix}\otimes\begin{bmatrix}
cos(x^d_2) \\
sin(x^d_2)
\end{bmatrix}\otimes...\otimes\begin{bmatrix}
cos(x^d_N) \\
sin(x^d_N)
\end{bmatrix}
\end{equation}
We map the \textit{N}-dimensional input vectors $x^d \in$ ${\rm I\!R}^N$ to a product state on \textit{N} qubits by using the feature map Eq. (2). The full quantum data is represented as tensor product $\phi^d=\otimes_{n=1}^N \phi_n^d$ \cite{huggins2018towards, gg}. Thus, the preparation of quantum state is efficient as it only needs single-qubit
rotations to encode each segment of classical dataset $n=\{1,2,...N\}$ in the amplitude of a qubit. Overall, there is no relevant cost for such encoding. Similar to classical dataset for binary classification, quantum data set for binary classification is denoted as a set $S_q=\{(\phi^d,y^d)\}_{d=1}^D$, where $y^d \in \{0,1\}$ are class labels for $2^N$-dimensional input vectors such that $\phi^d \in \mathbb{C}^{2^N}$. It can be easily checked that quantum data as a output of quantum circuit is in superposition state.  

\subsection{Quantum circuit classifier}

\begin{figure*}[!h]
\centering
\includegraphics[scale=0.6]{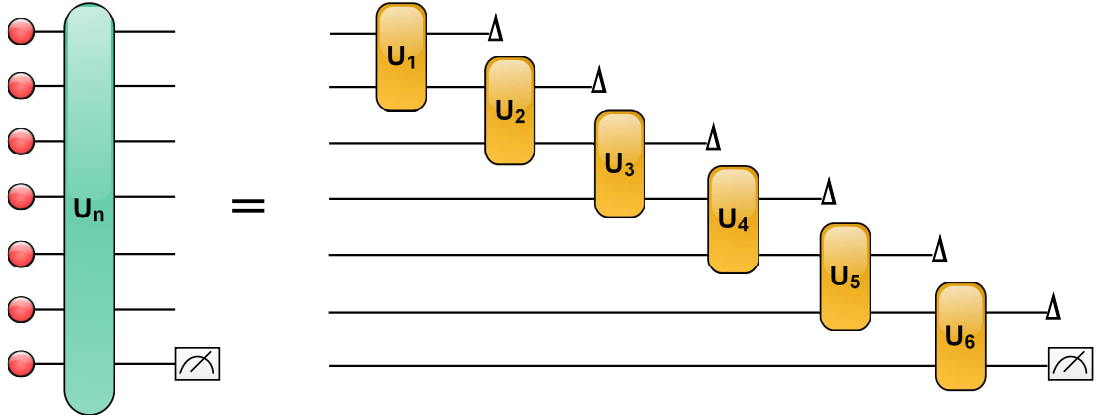}\hfil
\caption{Matrix product state quantum classifier}
\end{figure*}

We now discuss MPS quantum circuit classifier for classification of quantum data, which is made up of unitaries. We followed iterative approach by keeping positive trace values from \textit{N}-qubit
input space to output qubits.  We apply unitaries composed of single qubit rotations around \textit{y}-axis and CNOT gate to the input set and discard one of the qubits (unobserved) from each unitary. Therefore, we split the qubit into two parts for the next layer of the circuit. This process continues till the last qubit is left to be measured. It can be noted that at each stage of the circuit, we keep one of the qubits resulting from one of the unitary operations of the earlier stage and at last unitary transformation occurs on two qubits from another subpart of the circuit.

The unitary blocks in Fig 3 consists of input dataset with ancilla qubit which is initialized to zero. It can be easily traced out. Using ancilla qubit, we can execute large class of non-linear operations. In Fig 3, circles denote inputs prepared in a product state and triangles indicate unobserved qubits. At the end, when all unitary operations interpreted in the circuit have been executed, then one qubit is observed and used as the output qubit. The measurement on particular qubit is carried out by applying Pauli operators in particular direction. The output qubit determines the predicted value of the input, i.e. the class label values assigned. In order to calculate the most likely state of output qubit, the quantum circuit can be evaluated for number of iterations considering the same input to determine their probability distribution in the computational basis. The MPS quantum circuit for 7 qubits is represented by the Fig 3 consisting inputs represented by circles, unitary blocks $\{U_i\}^6_{i=1}$ and measurement operator on last qubit. Here, single-qubit rotations  in the \textit{y}-direction are followed by a CNOT gate and discard a qubit for next layer of the quantum circuit.

In order to assess the quality of actual and predicted values of dataset, we need to calculate the cost function. It measures the difference between actual and predicted values of dataset. It is given as:

\begin{equation}
J_{\theta}=\dfrac{1}{D}\sum_{d=1}^{D}(M_{\theta}(x^d)-y^d)^2
\end{equation}

where $x^d$ and $y^d$ are the input and class labels respectively, \textit{M} is qubit operator, $\theta$ represents the set of parameters to define the unitaries and \textit{D} is total number of data points. It calculates the average amount that the model's predictions differ from the actual values. The goal is to minimizing the cost function i.e. it must be close to zero. Although, there exists various procedures to carry out optimization. In order to optimize the large datasets, we have employed conjugate gradient (CG) method. It is iterative and effective method to optimize the results. But, it can be break down over multiple iterations when the function to be optimized is noisy. Alternatively, stochastic gradient descent method can be applied which is resilience to noise.

There are different parameters used to measure the performance of MPS quantum classifier such as accuracy (ACC), sensitivity (Sens), specificity (Spec) and gini coefficient. ACC is computed to measure the correctness of classifier,  Spec refers the the ability of classifier to identify the negative results. Sens defines the true positive rate i.e. correctly identified by the classifier, gini coefficient determines the inequality in the distribution and it should be between 0 and 1, where \textit{N} is the total number of data points, TP is True positive, TN defines True negative, FP and FN represent False positive and False negative respectively.

\begin{itemize}
\item Accuracy:
\begin{equation} 
ACC=  \dfrac{TP + TN}{N} \times 100 
\end{equation}
\item Sensitivity: \begin{equation} Sens=\frac{TP}{TP+FN} 
\end{equation}
\item Specificity: \begin{equation} Spec=\frac{TN}{TN+FP} 
\end{equation}
\end{itemize}

\section{Model development}
Here, we present the data preprocessing, encoding and managing of quantum data that are implemented on real-time quantum device ibmqx4, open source software and a programming language python 3.6.5 with installed Qiskit i.e. an open-source software development kit (SDK) for working with the IBM Q quantum processors.

\subsection{Data collection and preprocessing for the study}

Before, we develop the MPS based model for classification. Initially, we have collected the Iris sample datasets from UCI machine learning web portal. The Iris dataset consists of 150 sample examples in three varieties of Iris flowers (setosa, versicolor, virginica). We re-scaled the input vectors element-wise to lie in $[-\pi,\pi]$ and applied binary classification to class labels. After normalization process, we have formed pair wise combination of samples (Iris$_1$, Iris$_2$ and Iris$_3$) on the basis of classes such as Iris$_1$  consists data belong to class labels 1 and 2 (now encoded as 0 and 1). Similarly, Iris$_2$ and Iris$_3$ consist data belong to class labels 2, 3 and 1, 3 respectively.

\begin{figure*}[!h]
\centering
\includegraphics[scale=0.5]{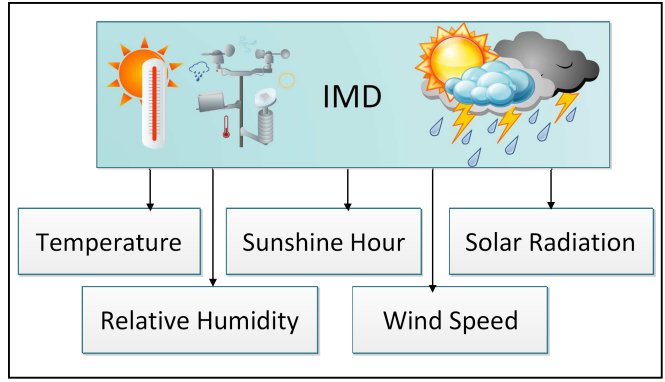}\hfil
\caption{Parameters of IMD weather dataset}
\end{figure*}

The climatic data of Patiala station has been retrieved from the India Meteorological Department (IMD), Pune. The station is located at the 30.33$^\circ$E latitude and 76.38$^\circ$S longitude. The study area includes Patiala station, located at the north part of Punjab (India). The elevation is 351m above sea level. The daily meteorological data for Patiala during (2014, 2015 and 2016) has been utilized. It consists following parameters: maximum and minimum air temperature (T$_{max}$) (T$_{min}$), relative humidity (R$_{H}$), wind speed (u$_{2}$), solar radiation (R$_{s}$), sunshine hours (I$_{s}$), evapotranspiration (ET$_{o}$), stand-deviation (SD), skewness (SK), kurtosis (K). The statistical parameters of meteorological variables at Patiala are given in Table 1. 
\begin{table*}[h]
\centering
\caption{Statistical parameters of available meteorological variables and ET$_{o}$ of Patiala station}
\begin{tabular}{ p{3.5 cm} | p{1.5 cm} p{1.5 cm} p{1.5 cm}  p{1.5 cm} p{1.5 cm}  p{1.5 cm}} 
	\hline 
	Parameters	&	Max	&	Min	&	Mean	&	SD	&	SK	&	K	\\
	\hline
	
	T$_{min}$ ($^\circ$C)	&	30.7	&	2.3	&	18.71	&	7.50	&	-0.27	&	-1.28	\\
	T$_{max}$ ($^\circ$C)	&	44.4	&	9.8	&	30.39	&	7.10	&	-0.47	&	-0.36	\\
	R$_{H}$(\%)	&	100	&	0	&	73.30	&	17.64	&	-0.77	&	-0.02	\\
	u$_{2}$ (km h$^{-1}$ day$^{-1}$)	&	16	&	0	&	3.23	&	2.18	&	1.46	&	3.16	\\
	I$_{s}$ (h)	&	12.2	&	0	&	6.24	&	3.53	&	-0.52	&	-0.98	\\
	R$_{s}$ (MJ m$^{-2}$ day$^{-2}$)	&	28.2	&	4.9	&	16.15	&	6.14	&	-0.01	&	-0.98	\\
	ET$_{o}$ (mm)	&	6	&	0	&	2.49	&	1.48	&	0.17	&	1.01	\\
	\hline
\end{tabular}
\end{table*}

In agriculture dataset, the ET$_o$ varies from 0 to 6. In order to perform binary classification, we have divided into three categories, the set \{0, 1\} comes under LOW , \{2, 3\} is MEDIUM and set \{4, 5, 6\} is categorized to HIGH as shown in Table 2. From this dataset we generated three binary classification tasks as C$_1$, C$_2$ and C$_3$. We re-scaled the input vectors element-wise to lie in $[-\pi,\pi]$ and applied binary classification to class labels. 

\begin{table*}[!ht]
\small
\centering
\caption{Performance comparison of MPS for each samples or Agriculture and Iris datasets}
\begin{tabular}{ p{1 cm} | p{3.5 cm} | p{1.2 cm} | p{4 cm} }
	\hline 
	{ET$_{o}$} 	& {Categories} 	&	{Classes} 	& 	{Samples} 	\\
	\hline
	
	0	&	(0-1)$\rightarrow$ LOW    &	C$_{1}$	& {Agri$_{1}$ (C$_{1}$ (0) \& C$_{2}$ (1))}	\\
	1	&	           	              &		    &                                       	\\
	\hline
	2	&	(2-3)$\rightarrow$ MEDIUM &	C$_{2}$	 & 	{Agri$_{2}$ (C$_{2}$ (0) \& C$_{3}$ (1))} \\
	3	&		                      &		     &	  	\\
	\hline
	4	&	(4-6)$\rightarrow$ HIGH	  &	C$_{3}$	  & {Agri$_{3}$ (C$_{1}$ (0) \& C$_{3}$(1))}  \\
	5	&		                      &	     	  &		\\
	6	&		                      &		      &	 	\\
	\hline
\end{tabular}
\end{table*}

After normalization process, we have formed pair wise combination of samples (Agri$_1$, Agri$_2$ and Agri$_3$) on the basis of classes such as Agri$_1$  consists data belong to class labels C$_1$ and C$_2$ (now encoded as 0 and 1). Similarly, Agri$_2$ and Agri$_3$ consist data belong to class labels C$_2$, C$_3$ and C$_1$, C$_3$ respectively. Further, each sample is divided into training and testing sets. The training set consists of 80\% and testing set consists of 20\% of the original dataset. Further, MPS quantum classifier is applied to the training dataset. It is repeated for number of iterations considering the same input. After achieving best accuracy, the trained model is applied to the testing dataset (unseen) and results are analyzed.

\begin{figure*}[!h]
\centering
\includegraphics[scale=0.5]{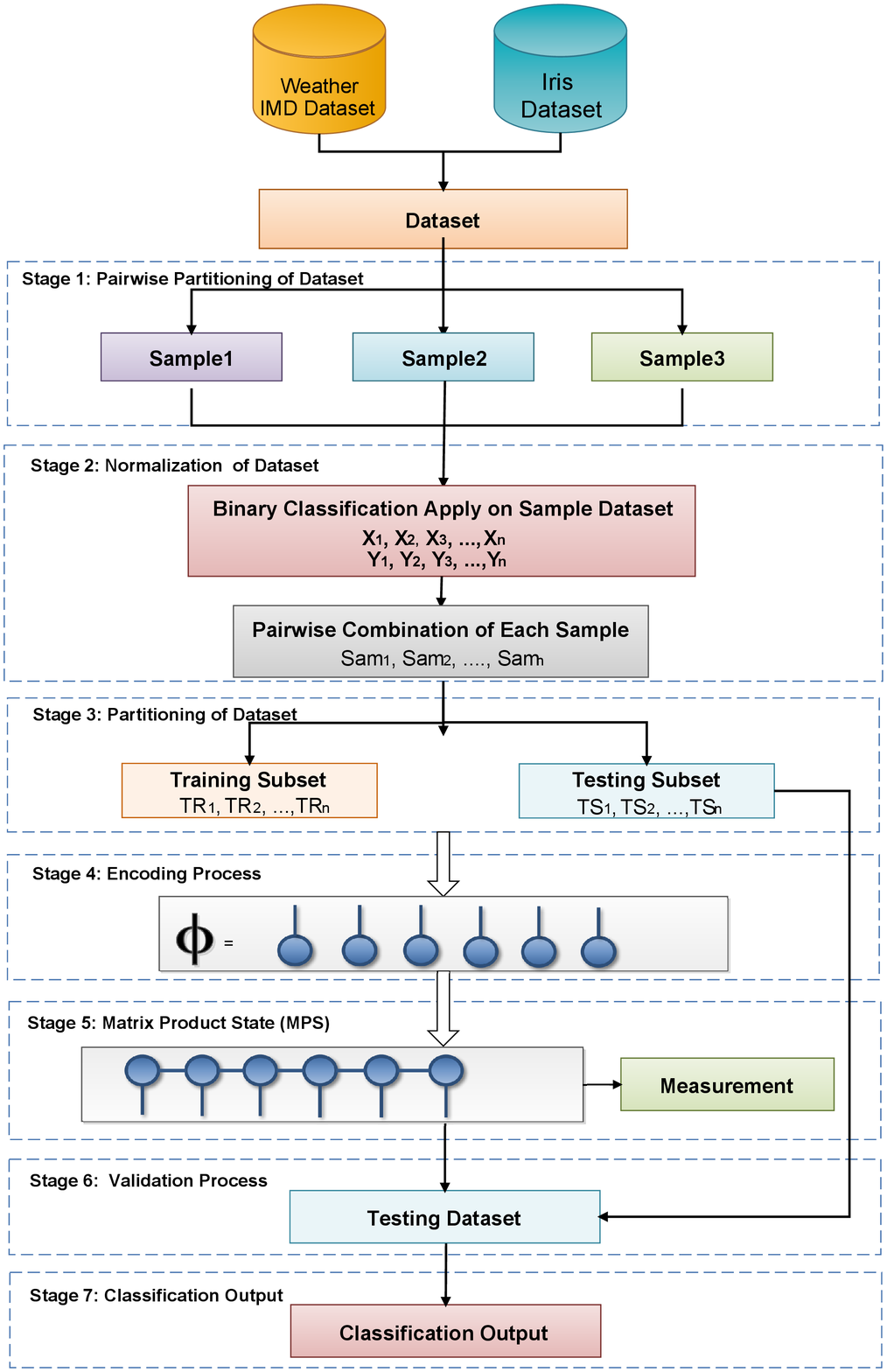}\hfil
\caption{Model development phases for classification}
\end{figure*}

\subsection{Development phases}

After partitioning of whole dataset in third phase, the training and testing sets are mapped into tensor network state by using Eq. (2). The input vectors are encoded into quantum state for classification of classical data on a quantum computer using Eq. (3). Finally, taken the tensor product of each input quantum state to form complete quantum data so that it can be ready to use in a quantum circuit using equation (4).  Fig 5 shows the seven stages of our methodology for model development.  

In fifth stage, qubit efficient MPS quantum classifier is trained using unitary parameters and qubit rotations in chosen direction. At the end, one or more qubits are measured using Pauli operators.

In final stage, we have determined the predicted value of the input, i.e. the class label values assigned for training set after assigning it to the quantum circuit. In order to calculate the most probable state of output qubit, the above stage is repeated for number of iterations considering the same input. Finally, we have the classification results for each sample. The experimental results are plotted in next section for each sample.

\subsection{Experimental results: Iris dataset}
In this Section, we have tested the ability of MPS quantum classifier to classify Iris dataset. In experimentation, we have given qubit rotations in \textit{y}-direction to have real values and parameterized the unitaries using ancilla qubit.  It is represented as four-qubit input gate consisting an ancilla qubit. It can be traced out in order to execute non-linear operations on dataset. In order to analyze the performance of MPS quantum classifier, we have divided the whole dataset into three samples (Iris$_1$, Iris$_2$, and Iris$_3$) on the basis of pairwise combination of class labels. Each sample consists 2/3 of the dataset. Further, we have  split each sample into a training set and testing test with 80:20 ratio to compute the accuracy. The classification of Iris dataset using MPS quantum classifier was done on the basis of accuracy and computed cost during the training and testing periods. On executing the MPS quantum classifier on each sample for classes 1 and 2, 2 and 3 as well as 1 and 3 of Iris dataset, it shows 85\% accuracy rate while differentiating classes 1 and 2, 80\% accuracy rate for classes 2 and 3, and 90\% accuracy rate differentiating classes 1 and 3. Therefore, it can be verified that accuracy results with largest values and cost with lowest values form the basis of higher model efficiency.

\begin{figure*}[!h]
\centering
\includegraphics[scale=0.7]{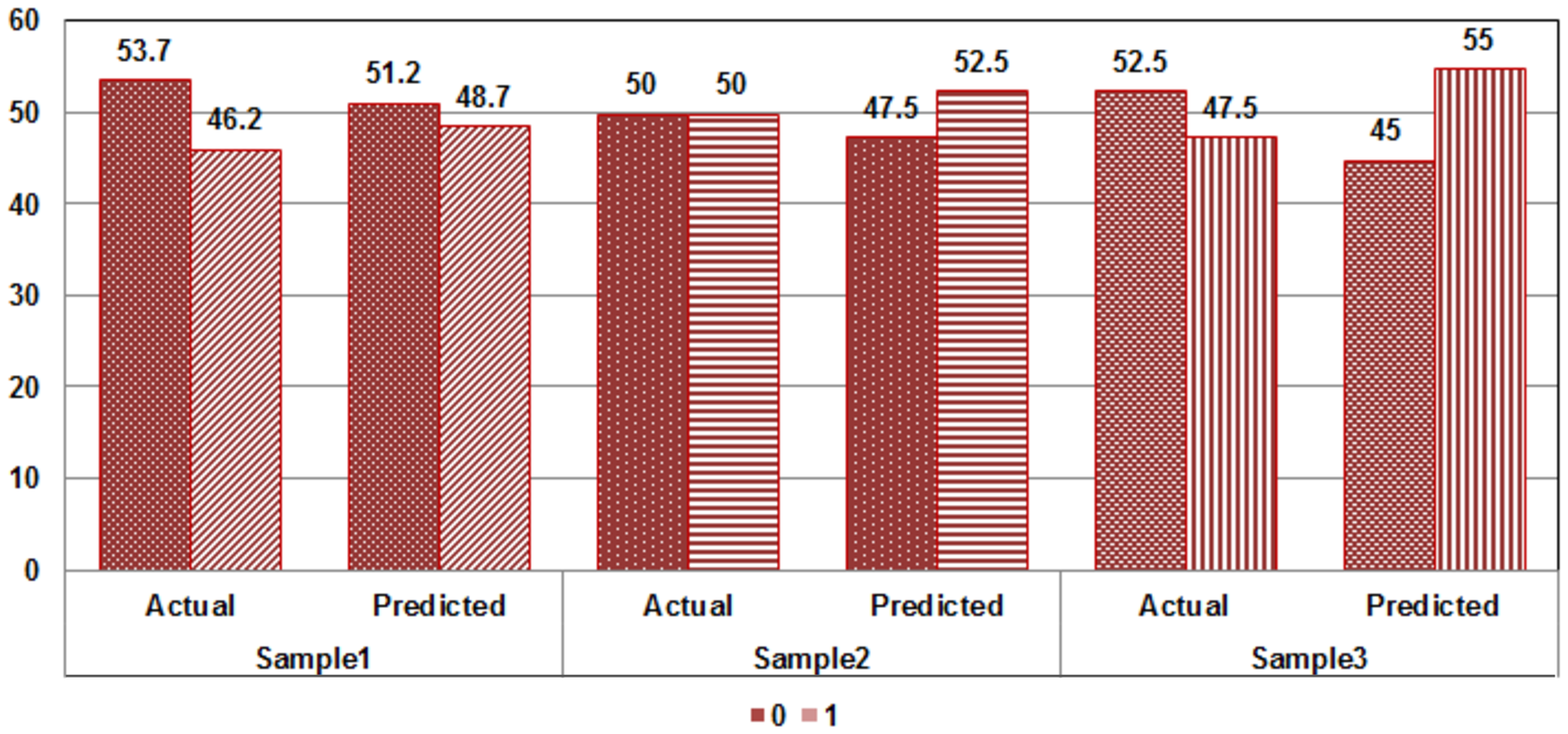}\hfil
\includegraphics[scale=0.7]{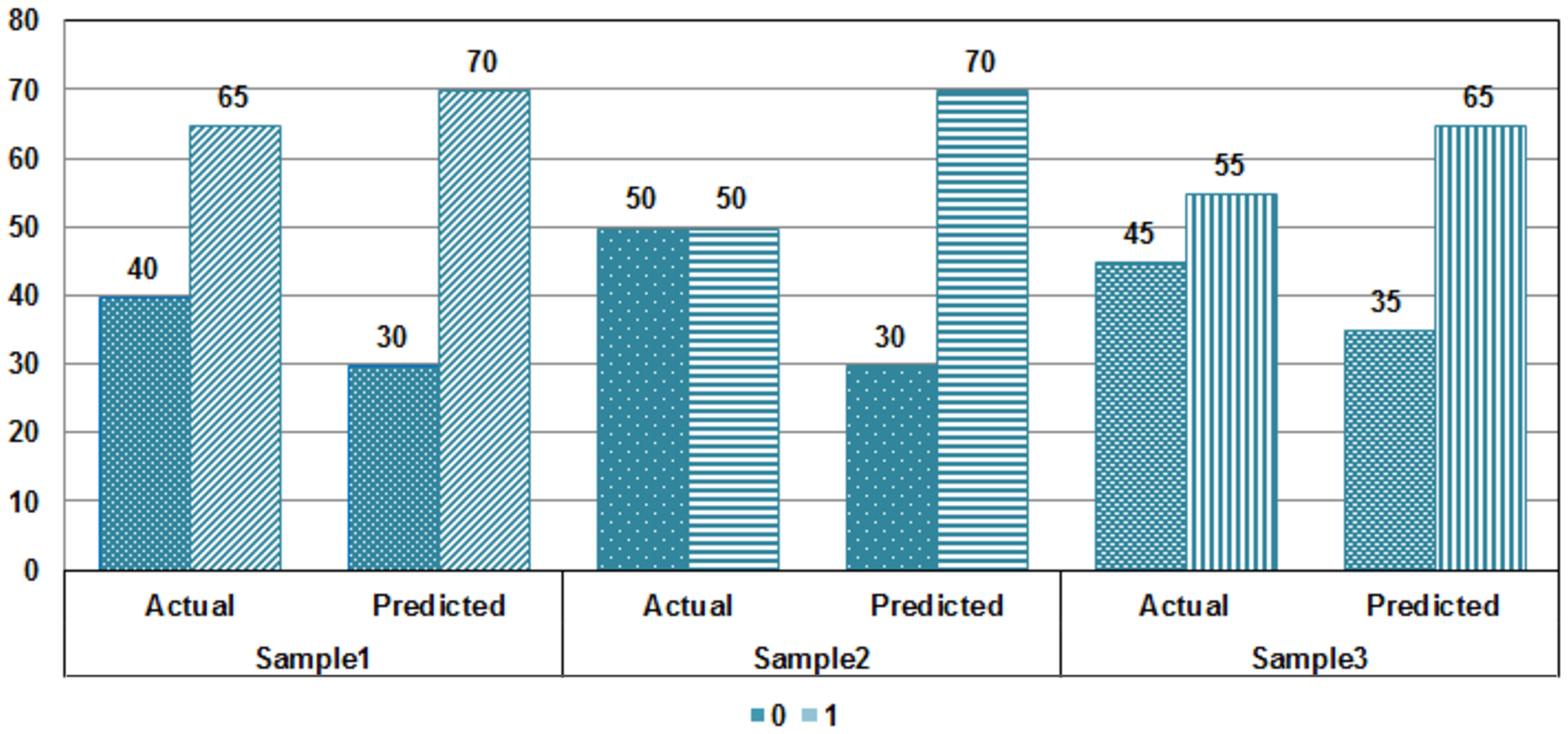}\hfil
\caption{Estimated IRIS$_{species}$ actual and predicted values (\%) for Training (in red), and Testing (in blue) datasets of Iris$_{1}$, Iris$_{2}$ and Iris$_{3}$}
\end{figure*}

\begin{table*}[!ht]
\centering
\caption{Performance comparison of MPS for each samples of Iris dataset}
\begin{tabular}{ p{1.1 cm} |p{0.8 cm} p{0.8 cm} p{0.8 cm}  p{0.8 cm} p{0.8 cm}| p{0.8 cm} p{0.8 cm} p{0.8 cm} p{0.8 cm} p{0.8 cm} } 
	\hline 
	Sample & \multicolumn{5}{c}{\textbf{Training}}  &  \multicolumn{5}{c}{\textbf{Testing}}   \\
	\hline
	&  Cost &  ACC & Spec & Sens & Gini &  Cost &  ACC & Spec & Sens & Gini  \\
	\hline
	Iris$_{1}$	&	0.11	& 88.75	& 0.86 & 0.90 & 0.80		 	& 0.15		& 85 & 0.85 & 0.83 &	0.67	    	\\
	Iris$_{2}$	&	0.16	& 83.75	& 0.82 & 0.84 &	0.80	 	& 0.2	&  80 & 0.71 & 1.0 & 0.50		   		\\
	Iris$_{3}$	&	0.05	& 95	& 0.90 & 1.0 & 0.92		& 0.1	& 90 & 0.84 & 1.0 & 0.77		   	\\
	\hline
\end{tabular}
\end{table*}

The output of MPS classifier is analyzed and compared for Iris different samples on the basis of parameters such as binary classification accuracy, cost, Spec, Sens and gini coefficient results as shown in Table 3. It can be easily checked that MPS quantum circuit gives results 88.75\% and 85\%, cost 0.11 and 0.15 for training and testing sets of Iris$_1$ respectively. Although, Iris$_2$ shows accuracy 83.75\% and 80\%, cost 0.16 and 0.2 for training and testing sets. Iris$_3$ results are slightly better than Iris$_1$ with accuracy 95\% and 90\% for training and testing sets. The MPS quantum classifier correctly classified the Iris dataset, and achieved cost function value of 0.05 and 0.1 for training and testing set of Iris$_3$ sample respectively (Eq. (5)).

Here, the experiments are performed with quantum circuits of \textit{N} = 4 qubits. We have produced datasets consisting 2500 quantum states for each of the classes $y \in\{1,2,3\}$. We have given each quantum state as an input into the quantum computer where the MPS quantum classifier is implemented. We have considered four-qubit input gates including ancilla qubit which is set to $\ket{0}$. The comparison between actual and predicted classification values in percentage for training and testing sets of Iris$_1$ (Sample1), Iris$_2$ (Sample2) and Iris$_3$ (Sample3) are shown in Fig 6.

\subsection{Experimental results: Agri dataset}

\begin{figure*}[!h]
	\centering
	\includegraphics[scale=0.7]{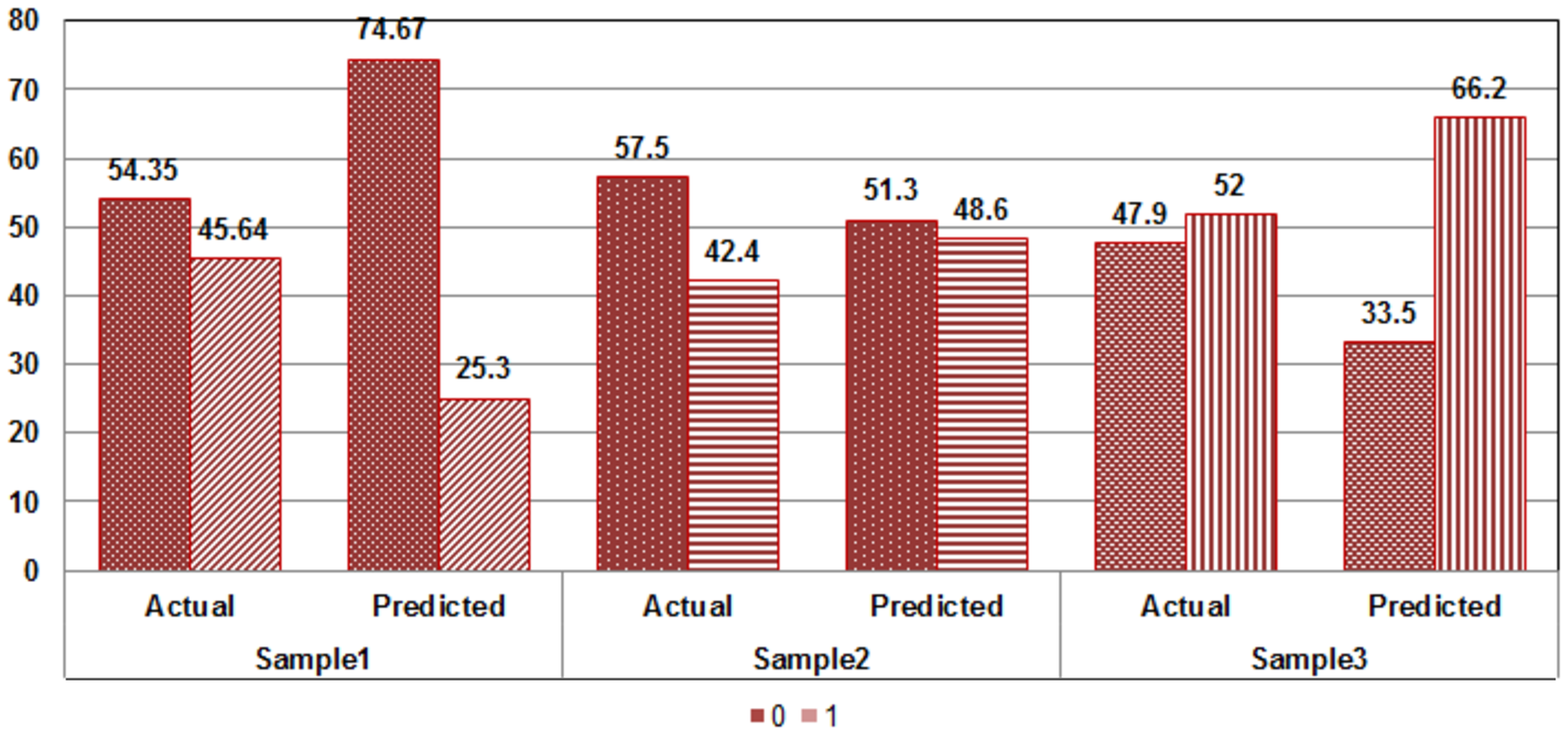}\hfil
	\includegraphics[scale=0.7]{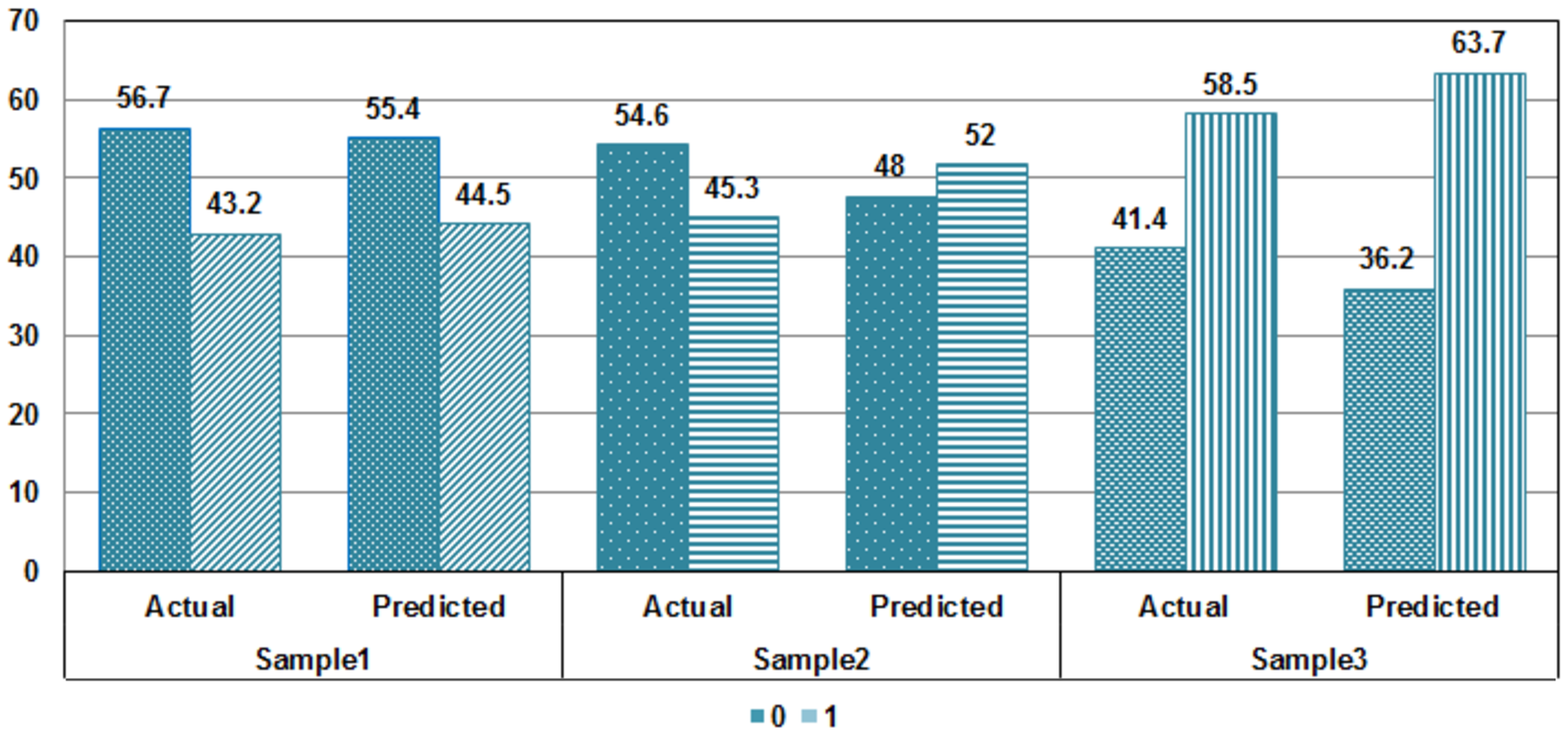}\hfil
	\caption{Estimated ET$_{o}$ Actual and Predicted values (\%) for Training (in red), and Testing (in blue) datasets of Agri$_{1}$, Agri$_{2}$ and Agri$_{3}$}
\end{figure*}

In this Section, the performance of proposed model is determined for agriculture domain using larger historical and streaming datasets. Each sample of Agri dataset consists statistical parameters (given in Table 1) formed by pairwise combination of classes. In order to test the ability of MPS classifier, we have trained it with training set as an input. The process is repeated considering the same input. Further, the testing set i.e. unseen data is given to quantum classifier. The performance of the proposed model during the training and testing period for each sample is given in Table 4. Compared with the each samples of training datasets, the accuracy of testing dataset of Agri$_{1}$ is just slightly greater. It can be easily checked that training accuracy of Agri$_2$ and Agri$_3$ samples is marginally higher than testing samples respectively. In case of training dataset of Agri$_1$ sample, the specificity is 0.98 approx i.e. MPS classifier identifies more negative results as compared to testing set 0.76. Therefore, the true positive value of training set is less than testing in case of Agri$_1$. The estimated  ET$_o$ actual and predicted values (in \%) for training and testing sets of Agri$_1$ (Sample1), Agri$_2$ (Sample2) and Agri$_3$ (Sample3) are plotted in Fig 7.

\begin{table*}[!ht]
\centering
\caption{Performance comparison of MPS for each samples of Agri dataset}
\begin{tabular}{ p{1.1 cm} |p{0.8 cm} p{0.8 cm} p{0.8 cm} p{0.8 cm}  p{0.8 cm}| p{0.8 cm} p{0.8 cm} p{0.8 cm} p{0.8 cm} p{0.8 cm}} 
	\hline 
	Sample & \multicolumn{5}{c}{\textbf{Training}}  &  \multicolumn{5}{c}{\textbf{Testing}}   \\
	\hline
	&  Cost &  ACC & Spec & Sens & Gini  & Cost &  ACC & Spec & Sens & Gini \\
	\hline
	Agri$_{1}$	&	0.20	& 79.03	& 0.98 & 0.72 &	0.52		& 0.19		& 80.65 & 0.76& 0.83 & 0.53		   		\\
	Agri$_{2}$	&	0.24	& 75.34	& 0.68 & 0.82 & 0.51	 	& 0.26	&  73.33 & 0.67 & 0.79 & 0.50		   		\\
	Agri$_{3}$	&	0.21	& 78.73	& 0.73 & 0.89 & 0.61	& 0.22	& 77.04 & 0.77 & 0.75 & 0.53		 		\\
	\hline
\end{tabular}
\end{table*}

\begin{figure*}[!h]
	\centering
	\includegraphics[scale=0.6]{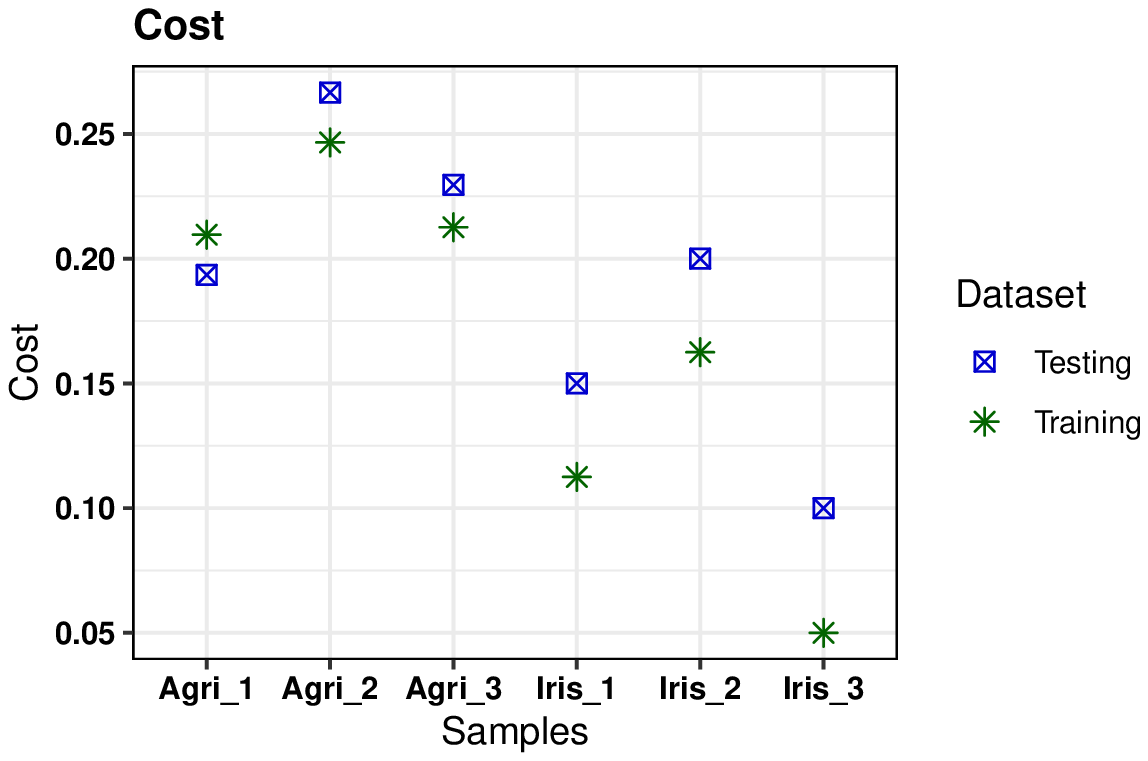}\hfil
	\includegraphics[scale=0.6]{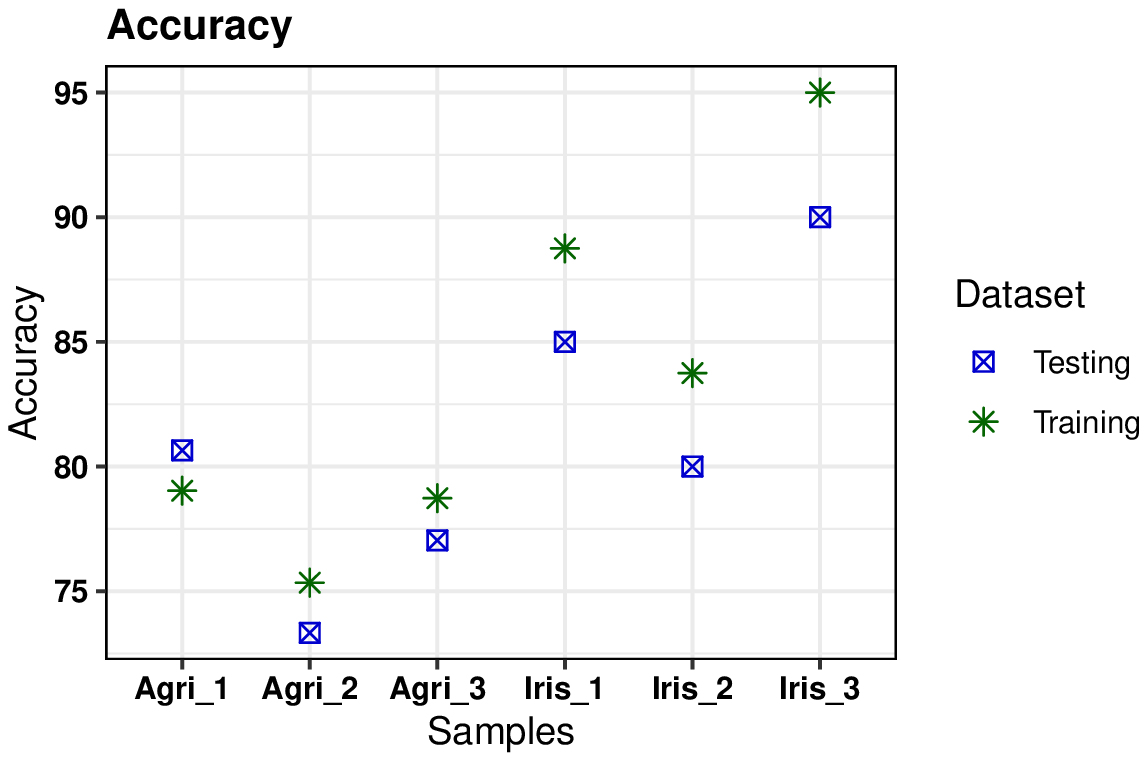}\hfil
	\includegraphics[scale=0.6]{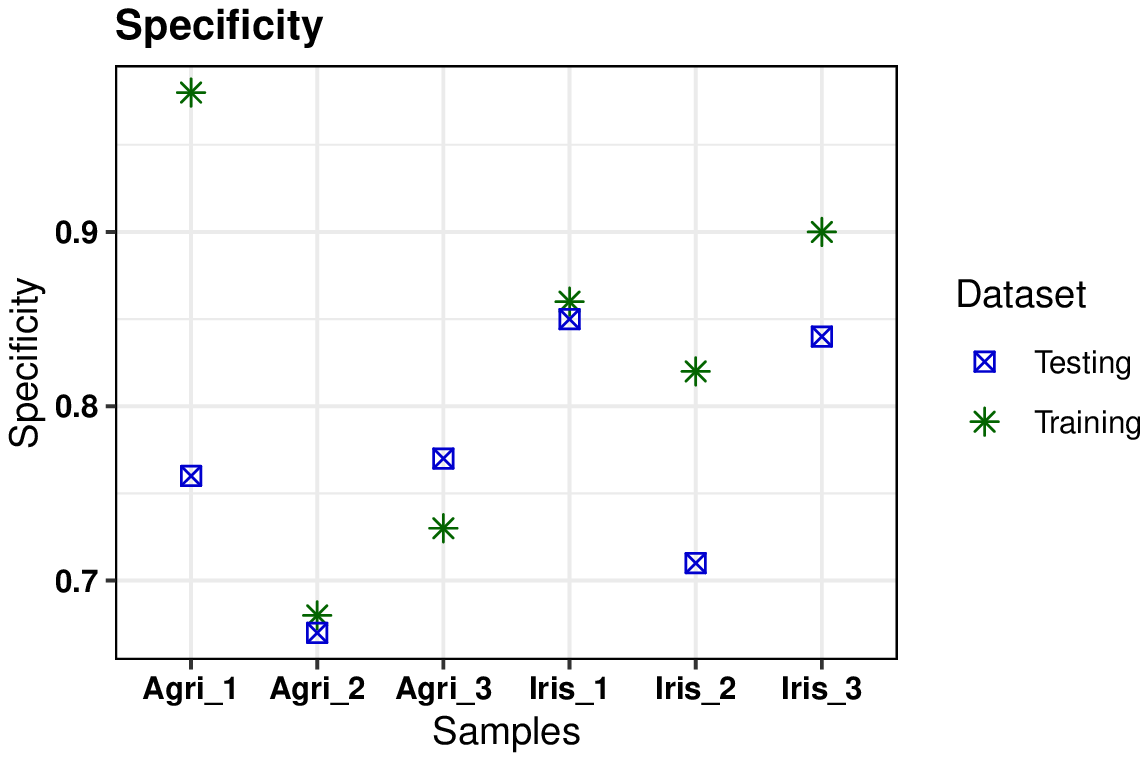}\hfil
	\includegraphics[scale=0.6]{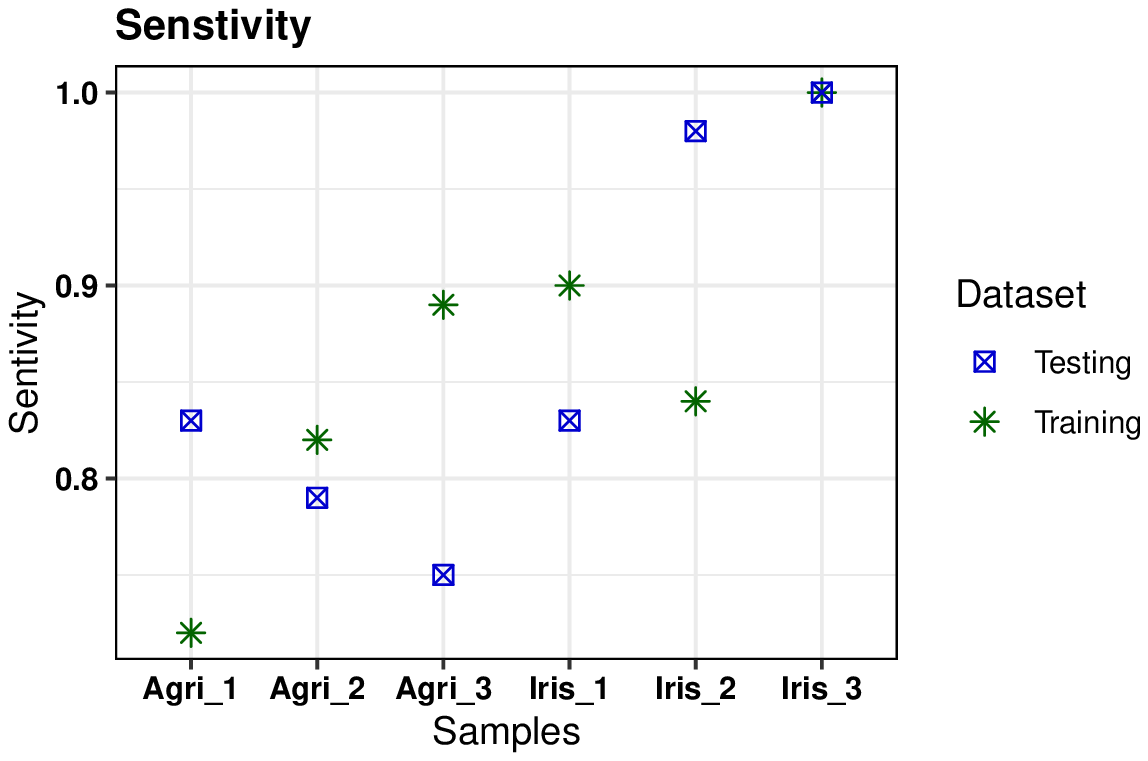}\hfil
	\includegraphics[scale=0.6]{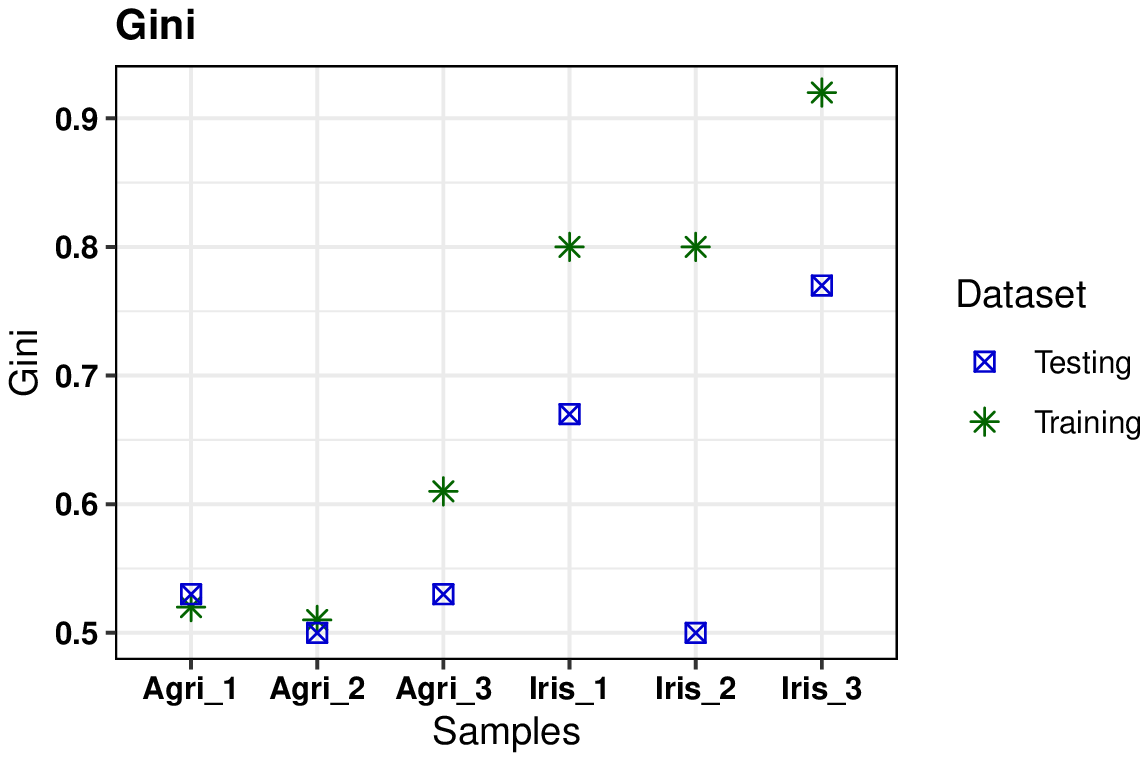}\hfil
	\caption{Comparison of forecasting ET$_{o}$ and Iris species results with MPS model (a) Cost, (b) ACC, (c) Spec, (d) Sens and (e) Gini}
\end{figure*}

The main advantage of MPS quantum classifier is that training can be implemented with high efficiency. The mapping of classical data into MPS form i.e. highly dimensional Hilbert space is really beneficial for generating high-order correlations between classes. The bond dimension of MPS manages the parameters of the machine learning model. It is easy to compute and can be selected adaptively. Usually, the bond dimension exists between (10-10,000) or more. It follows that larger dimension of bond results in higher accuracy. In fact, on selecting a extremely large bond dimension, the model can also result in overfitting, which is not in our case. MPS quantum classifier have been used to avoid overfitting as well as under-fitting, deal with corrected predictors and reduced the variance of the prediction error. It has been found to perform very effectively and efficiently handled big datasets. We believe that it can be adopted for many other machine learning tasks to scrutinize its power.
It has shown great learning capability for Iris species and ET$_{o}$ estimations in Agri dataset. Fig 8 describes the results of training and testing set accuracy, cost, Spec, Sens and gini coefficient for each sample and shows the consistency in accuracy of MPS quantum classifier.

\begin{figure*}[!h]
\centering
\includegraphics[scale=0.5]{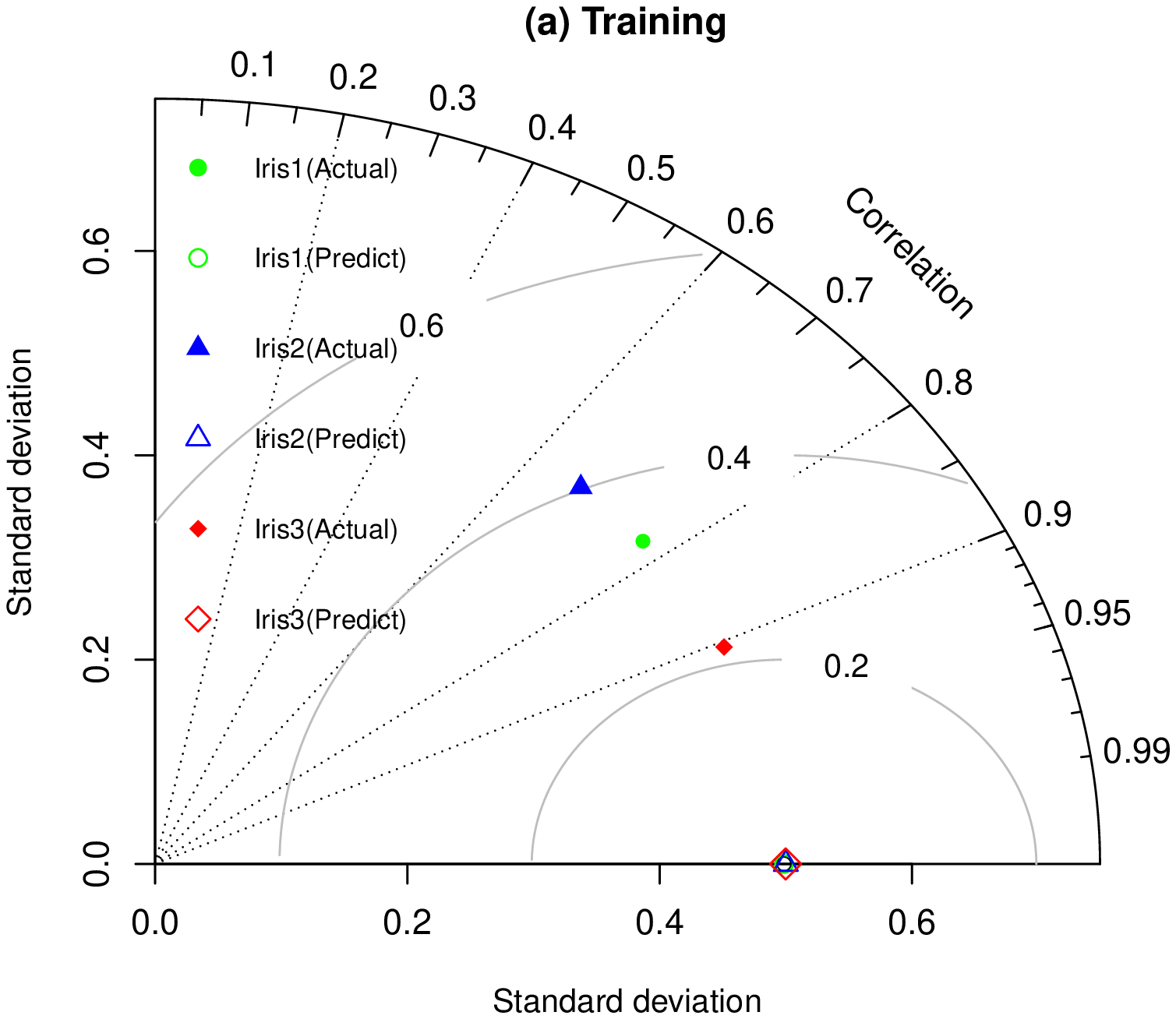}\hfil
\includegraphics[scale=0.5]{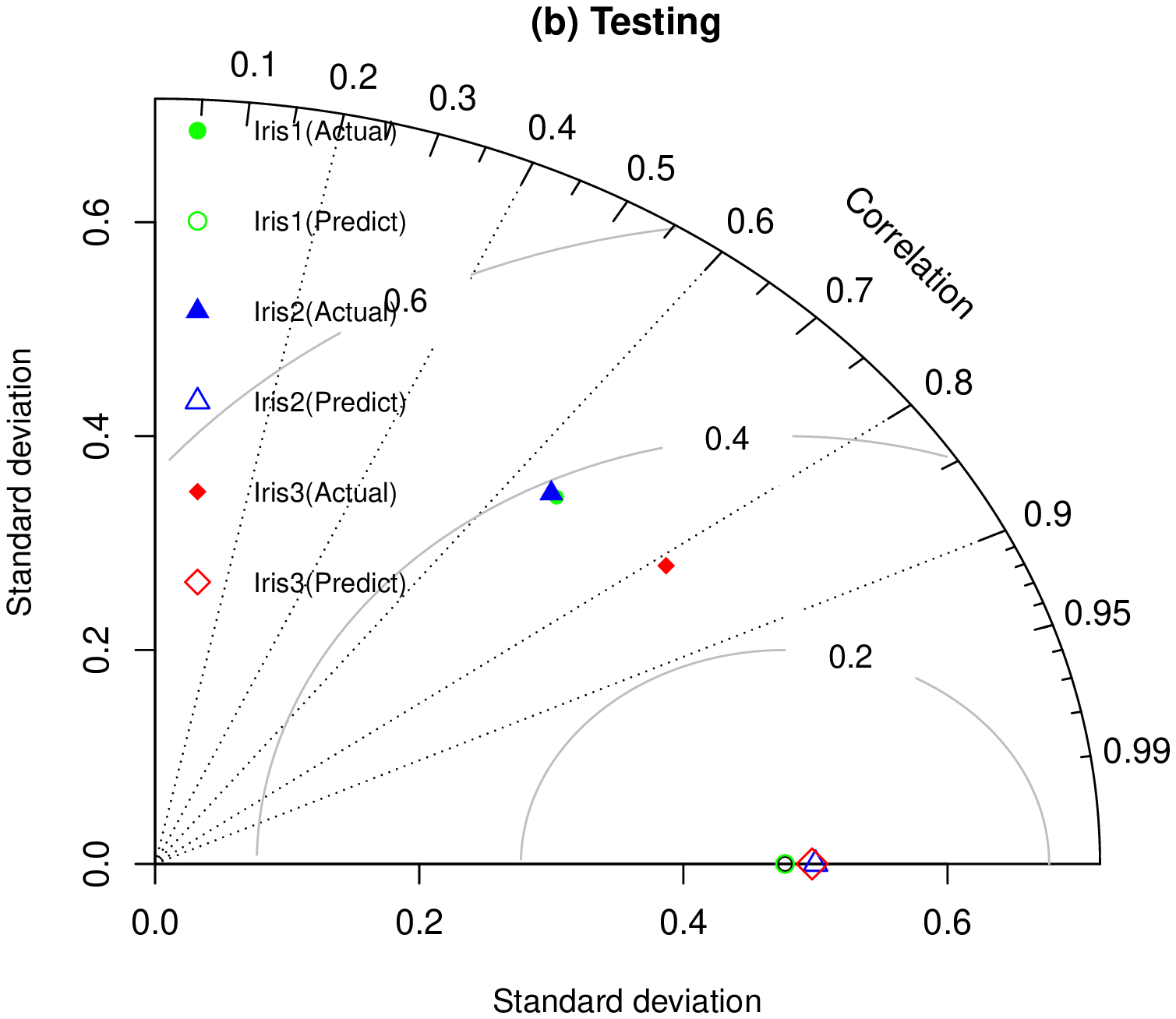}\hfil
\end{figure*}	
\begin{figure*}[!h]
\includegraphics[scale=0.55]{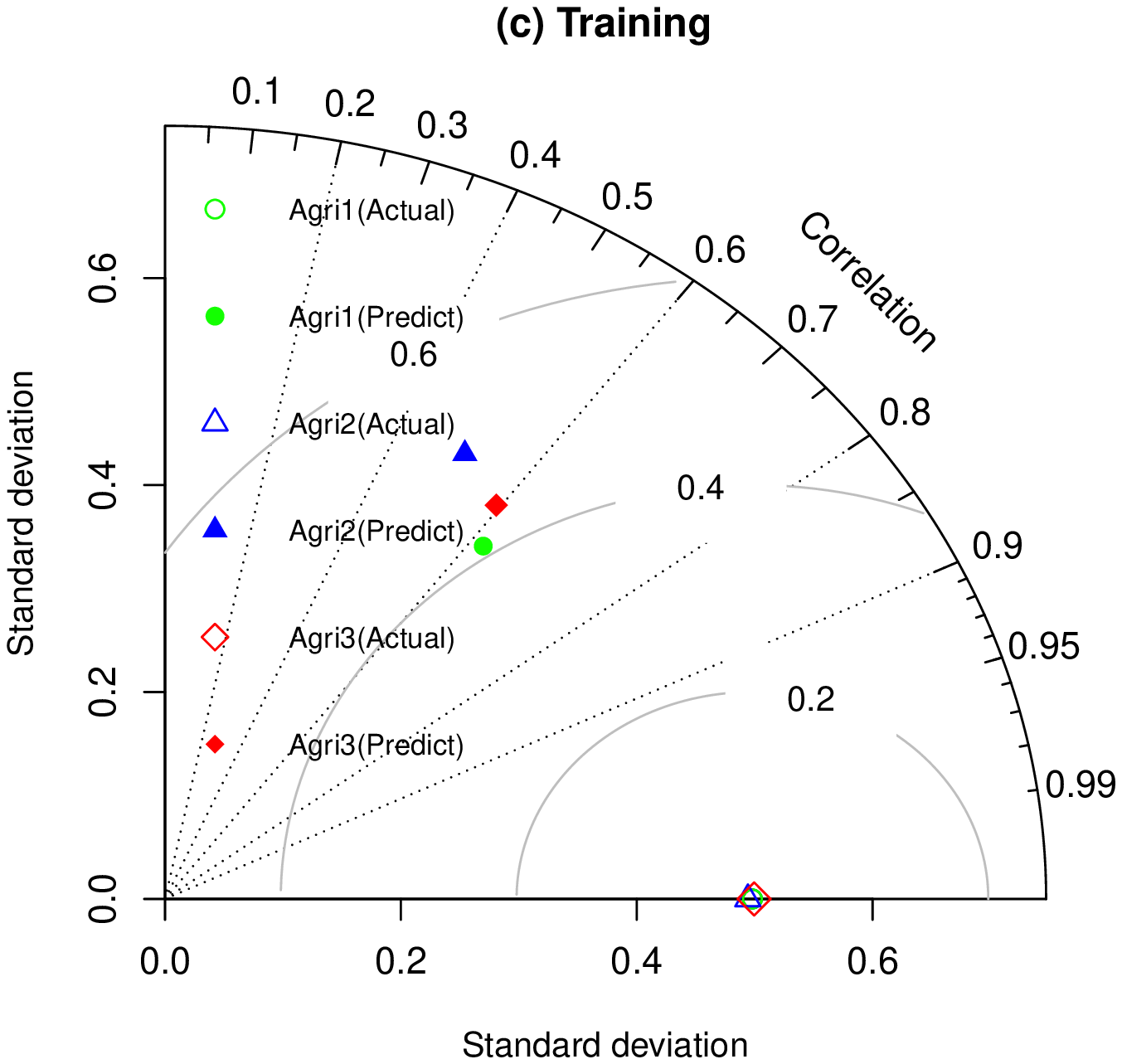}\hfil
\includegraphics[scale=0.55]{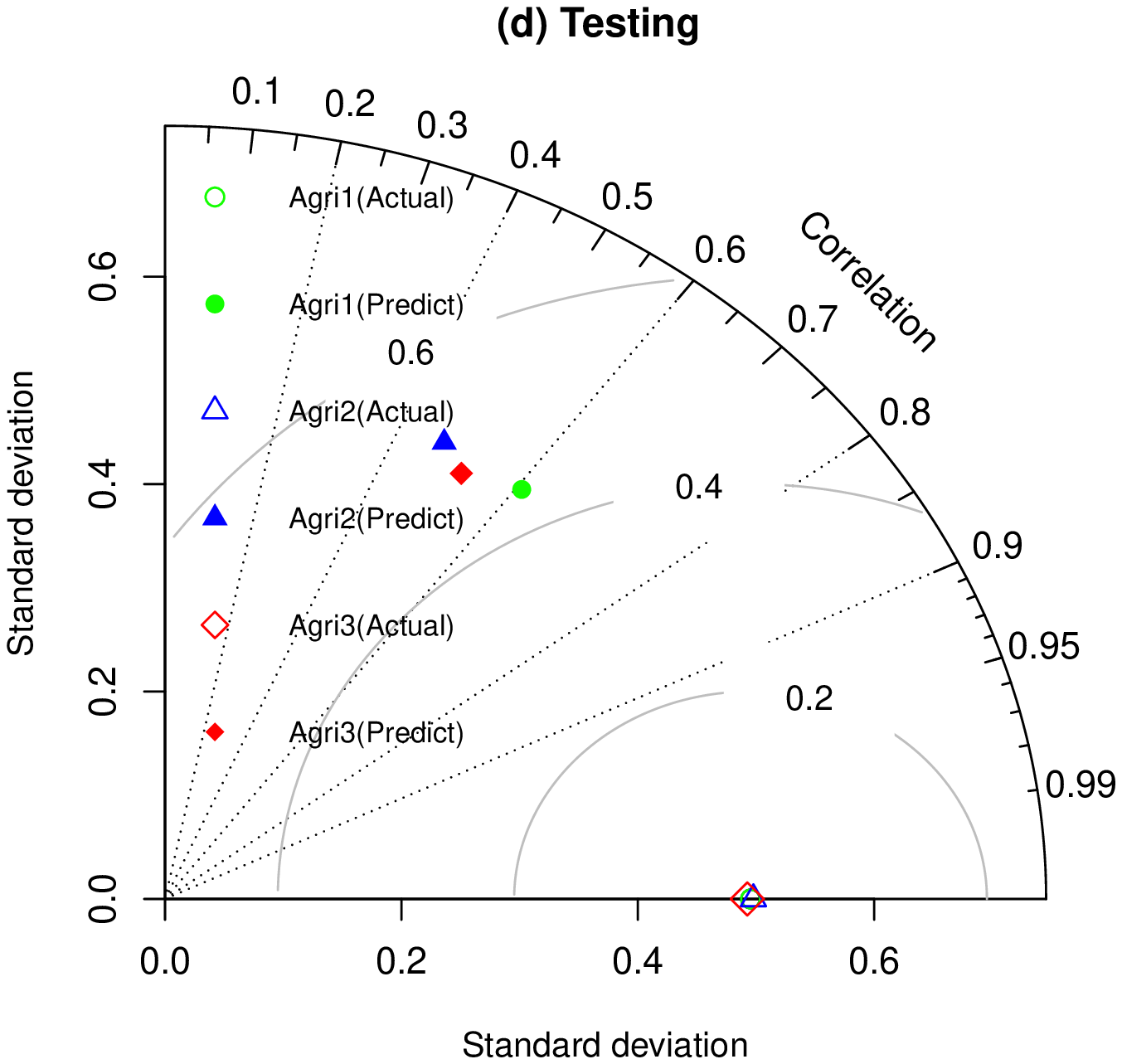}\hfil
\caption{Representation of degree of correspondence between each sample of Iris and Agri datasets (a) Training$_{Iris}$, (b) Testing$_{Iris}$, (c) Training$_{Agri}$, and (d) Testing$_{Agri}$ using Taylor diagrams}
\end{figure*}

The similarity between actual and predicted values is expressed on the basis of centered root-mean-square difference, correlation and their variations in amplitude i.e. standard deviation. We have used Taylor diagrams to graphically outline the degree of correspondence among values. It have been extensively used to investigate the performance of models to study aspects of climatic  environment. The colors indicate actual and predicted values of different samples for Iris and Agri datasets. Here, Iris1(actual), Iris1(predict) depict the actual and predicted values of training and testing sets of Iris dataset respectively. 

\section{Conclusion}

In this paper, we have illustrated that matrix product state quantum classifier can be used to classify quantum data efficiently. We have focused on MPS quantum circuit augmented with ancilla qubit that is implemented on quantum computer with restriction on qubit rotations to be real only. The key advantage of executing classification with MPS quantum circuit is that it can be executed efficiently with small number of qubits. The MPS quantum classifier has shown great learning capability for Iris and Agri datasets. Further, the different performance metrics of classification are analyzed for each sample and the degree of correspondence among values is shown. While the deployment of MPS quantum model to evaluate the other machine learning tasks is promising. In future, we will investigate the performance of other tensor networks for large real-time quantum data.

\section*{Acknowledgments}

Amandeep Singh Bhatia was supported by Maulana Azad National Fellowship (MANF), funded by Ministry of Minority Affairs, Government of India. Mandeep Kaur Saggi was supported by Council of Scientific \& Industrial Research (CSIR), funded by R\&D organization, India.

\end{document}